\documentclass[11pt]{article}
\usepackage[T1]{fontenc}
\usepackage[a4paper,margin=1in]{geometry}
\usepackage{amsmath,amssymb,amsthm}
\usepackage{graphicx}
\usepackage[hidelinks]{hyperref}

\newtheorem{theorem}{Theorem}
\newtheorem{lemma}{Lemma}
\newtheorem{proposition}{Proposition}

\newcommand{\B}{\mathcal{B}}
\newcommand{\cogirthratio}{\rho}
\newcommand{\cogirthratiowt}{\rho^{\mathrm{wt}}}

\title{Cogirth--Strength Ratios and Matroid Rank Reduction}
\author{Yu Cong\thanks{University of Electronic Science and Technology of China.
\href{mailto:yucong143@gmail.com}{\texttt{yucong143@gmail.com}}.}
\and Chao Xu\thanks{University of Electronic Science and Technology of China.
\href{mailto:thechaoxu@gmail.com}{\texttt{thechaoxu@gmail.com}}.}
\and Yajie Zhao\thanks{University of Electronic Science and Technology of China.
\href{mailto:yajiezhao8@gmail.com}{\texttt{yajiezhao8@gmail.com}}.}}
\date{}

\begin{document}
\maketitle

\begin{abstract}
In the rank-$p$ reduction problem, we are given a matroid with nonnegative element weights and seek a minimum-weight set of elements whose deletion lowers the rank by at least $p$.
For graphic matroids, this is the minimum $(p+1)$-cut problem.
Chekuri, Quanrud, and Xu developed a tree-packing algorithm for minimum $k$-cut, which relies on the fact that the minimum cut of a graph is at most twice its strength, the value of a maximum fractional packing of spanning trees.
We generalize this approach from graphs to matroids.
The matroid analogue of the minimum cut is the cogirth, the minimum weight of a cocircuit, and the analogue of graph strength is matroid strength, the value of a maximum fractional packing of bases.
We show that if every matroid in a deletion-closed class has cogirth at most $d$ times its strength, then for every $p$, the natural linear programming relaxation of rank-$p$ reduction is within a factor of $d$ of the optimum.
As a result, repeatedly deleting a minimum-weight cocircuit gives a $d$-approximation, and an optimal solution can be found deterministically in $n^{(2d-1)p+O(1)}$ time on $n$ elements.
For graphs, $d=2$ and our framework specializes to the tree-packing approach.

As applications, we bound the cogirth--strength ratio for two broad classes, and both bounds are tight.
For quasi-graphic matroids, which include graphic, bicircular, frame, and lifted-graphic matroids, the ratio is at most $2$ for all weights, giving a $2$-approximation and an exact $n^{3p+O(1)}$-time algorithm.
For $(k,\ell)$-count matroids of hypergraphs with $0\le \ell\le k$ and edge size at most $s$, the ratio is at most $ks$ under unit weights.
This gives a $ks$-approximation and an exact $n^{(2ks-1)p+O(1)}$-time algorithm for unweighted rank-$p$ reduction.
% The key tool for the latter is that, under unit weights, the ratio is subadditive under matroid union.
\end{abstract}

\section{Introduction}

Let $M$ be a matroid on a ground set $E$ with nonnegative element weights $w$.
In the \emph{rank-$p$ reduction} problem, studied by Joret and Vetta~\cite{JoretV15}, we seek a minimum-weight set of elements whose deletion lowers the rank of $M$ by at least $p$. We write $\lambda_{p,w}(M)$ for its optimum.
For the graphic matroid of a connected graph, deleting a set of edges lowers the rank by $p$ exactly when it leaves $p+1$ components, so rank-$p$ reduction is the minimum $(p+1)$-cut problem.
For $p=1$, it suffices to consider cocircuits, the inclusionwise minimal sets whose deletion lowers the rank.
The minimum weight of a cocircuit is the \emph{weighted cogirth} $\lambda_w(M):=\lambda_{1,w}(M)$, the matroid analogue of the minimum cut.

A set lowers the rank by at least $p$ if and only if it contains at least $p$ elements of every base.
Writing $\B(M)$ for the set of bases of $M$ and $x(S):=\sum_{e\in S}x_e$, this leads to the \emph{base-hitting LP}
\begin{equation}\label{eq:base-hitting}
\lambda_{p,w}^*(M):=\min\Bigl\{\sum_{e\in E}w_ex_e \;\Bigm|\; x\in[0,1]^E,\; x(B)\ge p\text{ for }B\in\B(M)\Bigr\},
\end{equation}
which generalizes the LP formulation of minimum $k$-cut~\cite{chekuri_lp_2020}. Its dual is
\begin{equation}\label{eq:base-packing}
\max\Bigl\{ p\sum_{B\in\B(M)}y_B-\sum_{e\in E}z_e \;\Bigm|\; y,z\ge0,\; \sum_{B\ni e}y_B-z_e\le w_e\text{ for }e\in E\Bigr\}.
\end{equation}
For $p=1$, the upper bounds $x_e\le1$ are redundant, so we may take $z=0$, and \eqref{eq:base-packing} asks for a maximum fractional packing of bases in which each element $e$ has capacity $w_e$.
Its optimum is the \emph{weighted strength} $\sigma_w(M)$~\cite{catlin_fractional_1992,Karger98}.
Since such a packing is also feasible for \eqref{eq:base-packing} for every $p$, we have $\lambda_{p,w}(M)\ge\lambda_{p,w}^*(M)\ge p\,\sigma_w(M)$.

In this paper we focus on the \emph{cogirth--strength ratio} $\lambda_w(M)/\sigma_w(M)$, which is the integrality gap of \eqref{eq:base-hitting} for $p=1$.
For unit weights we drop the subscript $w$ and write $\cogirthratio(M):=\lambda(M)/\sigma(M)$.
In graphs, the theorem of Nash-Williams and Tutte~\cite{Nash-Williams_1961,Tutte_1961} implies that the minimum cut is at most twice the strength, and this fact underlies the tree-packing approach to minimum $k$-cut~\cite{chekuri_lp_2020}.
For general matroids, the ratio can be arbitrarily large.\footnote{For instance, the uniform matroid $U_{r,2r}$ has ratio $(r+1)/2$.}
We ask which matroids have bounded cogirth--strength ratio, and what a bounded ratio implies for rank-$p$ reduction when $p>1$.

\paragraph{Previous work.}
As a special case of rank reduction, the minimum $k$-cut problem has been studied extensively~\cite{goldschmidt_polynomial_1994,saran_finding_1995,thorup_minimum_2008,GuptaLL18,Li_2019,gupta_optimal_2022}.
Two lines of work are most relevant to us.
The first is the LP-based tree-packing approach of Chekuri, Quanrud, and Xu~\cite{chekuri_lp_2020}, which bounds the integrality gap of the spanning-tree relaxation of $k$-cut and derives an exact enumeration algorithm from an optimal fractional tree packing.
Our framework generalizes this approach from spanning trees to bases of matroids.
The second is the series of works built on Karger's random contraction algorithm~\cite{karger_global_1993,karger_new_1996,gupta_optimal_2022}, which very recently inspired a random deletion algorithm of Chandrasekaran, Chekuri, and Zhu~\cite{chandrasekaran_random_2026}: it enumerates all $\alpha$-approximate minimum circuits of an $m$-element regular matroid in randomized $m^{O(\alpha)}$ time.

Beyond graphic matroids, Joret and Vetta~\cite{JoretV15} showed that rank reduction is polynomial-time solvable for partition matroids but NP-hard for intersections of two partition matroids.
Other works have focused on cut problems for specific matroid classes.
For a hypergraph $H$ and integers $k,\ell$, the \emph{$(k,\ell)$-count matroid} $M_{k,\ell}(H)$ has as independent sets the edge sets $I$ such that $|J|\le k|V(J)|-\ell$ for every nonempty $J\subseteq I$, where $V(J)$ is the set of vertices met by $J$.
Narayanan, Roy, and Patkar~\cite{narayanan_approximation_1996} studied approximation algorithms for hypergraph cut problems, and Fukunaga~\cite{fukunaga_computing_2013} gave an exact algorithm for hypergraph $k$-cut when both $k$ and the maximum hyperedge size are fixed.
Király~\cite{Kiraly09} proved that the unweighted cogirth of $M_{k,k}(H)$ can be computed in polynomial time without any bound on the hyperedge size.
Cong, Xu, and Zhao~\cite{cong_destroying_2026} recently proved that computing the unweighted cogirth of the two-dimensional rigidity matroid of a simple graph, which is its $(2,3)$-count matroid, is NP-hard.
Geelen and Kapadia~\cite{geelen_computing_2018} gave randomized fixed-parameter tractable algorithms for computing the girth and the cogirth of rank-$t$ perturbed graphic matroids, and Cong, Xu, and Zhao~\cite{cong_derandomizing_2026} derandomized the cogirth algorithm.
The relationship between cogirth and base packing has also been studied independently: Bailey, Newman, and Stevens~\cite{BaileyNS14} investigated matroids whose cogirth equals their integral base-packing number, and noted the difficulty of recognizing this equality in polynomial time for general matroids.

These results show that rank reduction behaves very differently across matroid classes, and most algorithms exploit structure specific to one class.
It is natural to ask whether a single condition yields approximation guarantees and exact algorithms for many classes at once.
We show that a bounded cogirth--strength ratio is such a condition.

\paragraph{Our contributions.}
Throughout, matroid $M$ has $n$ elements and is given by an independence oracle.
Our first result shows that the integrality gap for $p=1$ controls the integrality gap for every $p$.
If $\lambda_w(N)\le d\,\sigma_w(N)$ for every positive-rank restriction $N$ of $M$, then $\lambda_{p,w}(M)\le d\,\lambda_{p,w}^*(M)$ for every $p$, and deleting a minimum-weight cocircuit $p$ times attains this bound (\autoref{thm:rank-reduction-gap}).
Adapting the enumeration algorithm of Chekuri, Quanrud, and Xu~\cite{chekuri_lp_2020}, with bases in place of spanning trees and flats in place of graph components, we compute an optimal solution deterministically in $n^{(2d-1)p+O(1)}$ time, and implement the greedy $d$-approximation in $p\,n^{2d-1+O(1)}$ time (\autoref{thm:lp-enumeration}).
Neither algorithm needs to know $d$.
As a by-product, we decide in polynomial time whether the cogirth of a matroid equals its integral base-packing number, which resolves the recognition problem noted by Bailey, Newman, and Stevens~\cite{BaileyNS14}.

We then bound the cogirth--strength ratio for two deletion-closed classes.
\begin{itemize}
\item Quasi-graphic matroids, which include graphic, bicircular, frame, and lifted-graphic matroids~\cite{bowler_describing_2025}, have weighted ratio at most $2$ (\autoref{thm:quasi-graphic-ratio}). This gives a $2$-approximation and an exact $n^{3p+O(1)}$-time algorithm for weighted rank-$p$ reduction.
\item For $0\le \ell\le k$, the count matroid $M_{k,\ell}(H)$ of a hypergraph with edges of size at most $s$ has ratio at most $ks$ under unit weights (\autoref{thm:count-unit-bound}). This gives a $ks$-approximation and an exact $n^{(2ks-1)p+O(1)}$-time algorithm for unweighted rank-$p$ reduction.
\end{itemize}
Both bounds are tight. The one for count matroids follows from the subadditivity of the unweighted ratio under matroid union (\autoref{thm:ratio-subadditivity-under-union}).

\section{Preliminaries}
We assume familiarity with matroid theory and follow the notation of~\cite{Oxley06}.
For a matroid $M$ on ground set $E=E(M)$, we write $r_M$ for its rank function, $r(M):=r_M(E)$ for its rank, and $\B(M)$ for its set of bases.
For $X\subseteq E$, we write $M\setminus X$, $M/X$, and $M|X:=M\setminus(E\setminus X)$ for the matroids obtained by deleting $X$, contracting $X$, and restricting to $X$, respectively.
The closure $\operatorname{cl}_M(X)$ of $X\subseteq E$ is the set of elements whose addition to $X$ does not increase the rank.
A flat is a set equal to its closure, and a hyperplane is a flat of rank $r(M)-1$.
A cocircuit is a minimal set whose deletion decreases the rank.
Cocircuits are exactly the complements of hyperplanes.
% An independence oracle answers whether a given set is independent, and we count each query as unit time.
For $x\in\mathbb R^E$ and $S\subseteq E$, we write $x(S):=\sum_{e\in S}x_e$.

Let $w:E\to\mathbb R_{\ge0}$ be weights; on minors of $M$, $w$ denotes the restricted weights. The weighted cogirth $\lambda_w(M)$ is the minimum weight of a cocircuit.
The weighted strength $\sigma_w(M)$ is the minimum of $\frac{w(E\setminus F)}{r(M)-r_M(F)}$ over all flats $F$ of rank less than $r(M)$; by the base-packing theorem, it equals the optimum of~\eqref{eq:base-packing} for $p=1$ (see Section~\ref{sec:lp-reduction}).
Unless stated otherwise, all matroids considered are loopless and of positive rank.
For unit weights we drop the subscript $w$, and the cogirth--strength ratio is $\cogirthratio(M):=\lambda(M)/\sigma(M)$.
For weights with $\sigma_w(M)>0$, we write $\cogirthratiowt_w(M):=\lambda_w(M)/\sigma_w(M)$, and the \emph{weighted cogirth--strength ratio} $\cogirthratiowt(M)$ is the supremum of $\cogirthratiowt_w(M)$ over all such weights.

Let $M_1,M_2$ be matroids on disjoint ground sets $E_1,E_2$, respectively.
Their direct sum $M=M_1\oplus M_2$ is the matroid on $E_1\cup E_2$ with independent sets $\mathcal I(M)=\{I_1\cup I_2: I_1\in\mathcal I(M_1),\ I_2\in\mathcal I(M_2)\}$.
\begin{lemma}
\label{lem:direct-sum}
Let $M=M_1\oplus\cdots\oplus M_m$ and let $w:E(M)\to\mathbb R_{\ge 0}$. Then $\lambda_w(M)=\min_i\lambda_w(M_i)$ and $\sigma_w(M)=\min_i\sigma_w(M_i)$.
\end{lemma}

\begin{proof}
The cocircuits of $M$ are exactly the cocircuits of the summands $M_i$, giving $\lambda_w(M)=\min_i\lambda_w(M_i)$.

Every flat of $M$ has the form $F=\bigcup_i F_i$, where $F_i$ is a flat of $M_i$,
and $r_M(X)=\sum_i r_{M_i}(X\cap E(M_i))$ for all $X\subseteq E(M)$.
If $r_M(F)<r(M)$, then
\[
\begin{aligned}
\frac{w(E(M)\setminus F)}{r(M)-r_M(F)}
=\frac{\sum_{i}w(E(M_i)\setminus F_i)}{\sum_{i}\bigl(r(M_i)-r_{M_i}(F_i)\bigr)}
&\ge\min_{i:\,F_i\ne E(M_i)}\frac{w(E(M_i)\setminus F_i)}{r(M_i)-r_{M_i}(F_i)}\\
&\ge\min_i\sigma_w(M_i).
\end{aligned}
\]
Equality holds when $F_i$ attains $\sigma_w(M_i)$ for a minimizing $i$ and $F_j=E(M_j)$ for $j\ne i$.
Hence $\sigma_w(M)=\min_i\sigma_w(M_i)$.
\end{proof}
In particular, $\cogirthratiowt_w(M)\le\max_i\cogirthratiowt_w(M_i)$ whenever $\sigma_w(M)>0$,
since any $M_i$ attaining $\sigma_w(M)$ has cogirth at least $\lambda_w(M)$.

\section{Rank-\texorpdfstring{$p$}{p} reduction}

Throughout this section, let $p\in\mathbb Z_+$, and let $M$ be an $n$-element matroid of rank at least $p$ with element weights $w:E(M)\to\mathbb R_{\ge0}$.
Rank-$p$ reduction asks for a minimum-weight set whose deletion lowers the rank by at least $p$.
A set $S\subseteq E(M)$ is a \emph{$p$-cocycle} if $E(M)\setminus S$ is a flat and $r_M(E(M)\setminus S)\le r(M)-p$.

Every rank-$p$ reduction set $R$ contains the $p$-cocycle $E(M)\setminus\operatorname{cl}_M(E(M)\setminus R)$.
Hence rank-$p$ reduction is the minimum-weight $p$-cocycle problem. We write $\lambda_{p,w}(M)$ for its optimum.

\subsection{The base-hitting LP and its gap}
\label{sec:lp-reduction}

We bound the integrality gap of~\eqref{eq:base-hitting} by following the greedy algorithm that repeatedly deletes a minimum-weight cocircuit.
Recall that $\lambda_{1,w}^*(M)=\sigma_w(M)$~\cite[Section~42.4]{schrijver_combinatorial_2003}.
Deleting a cocircuit $C$ lowers the rank by exactly one, so $C$ together with any rank-$(p-1)$ reducing set of $M\setminus C$ is a rank-$p$ reducing set of $M$.
The next lemma is the fractional counterpart: deleting $C$ and lowering the target by one decreases the LP optimum by at least $\sigma_w(M)$.

\begin{lemma}
\label{lem:cocircuit-lp-drop}
Let $C$ be a cocircuit of $M$. Then $\lambda_{1,w}^*(M)+\lambda_{p-1,w}^*(M\setminus C)\le\lambda_{p,w}^*(M)$,
with the convention $\lambda_{0,w}^*\equiv0$.
\end{lemma}

\begin{proof}
The case $p=1$ is immediate, so assume $p\ge 2$.
Put $E:=E(M)$, let $x$ be any feasible solution to the rank-$p$ LP~\eqref{eq:base-hitting} on $M$, and set $a:=\min_{e\in C}x_e\in[0,1]$.
Extend $x':=\frac{p-1}{p-a}\,x|_{E\setminus C}$ by zero on $C$.
Since $0\le a\le 1$ and $p\ge 2$, we have $0\le x'\le x\le 1$ coordinatewise, so both $x'$ and $x-x'$ lie in $[0,1]^E$.

Since $C$ is a cocircuit, $E\setminus C$ is a hyperplane, so every base $B'$ of $M\setminus C$ extends to a base of $M$ by adding any $e\in C$.
Choosing $e\in C$ with $x_e=a$ gives $x(B')\ge p-a$ and hence $x'(B')\ge p-1$, so $x'|_{E\setminus C}$ is feasible for the rank-$(p-1)$ LP on $M\setminus C$.

Moreover, every base $B$ of $M$ meets the cocircuit $C$, so $x(B\cap C)\ge a$. Together with $x(B)\ge p$ and $1-a\ge 0$, this gives
\[
\begin{aligned}
(x-x')(B)
&=\frac{1-a}{p-a}\,x(B\setminus C)+x(B\cap C)\\
&=\frac{(1-a)x(B)+(p-1)x(B\cap C)}{p-a}
\ge\frac{(1-a)p+(p-1)a}{p-a}=1,
\end{aligned}
\]
so $x-x'$ is feasible for the rank-one LP on $M$.
Since $x'$ vanishes on $C$, the objective values of $x-x'$ and $x'|_{E\setminus C}$ sum to $\sum_{e\in E}w_ex_e$, giving $\lambda_{1,w}^*(M)+\lambda_{p-1,w}^*(M\setminus C)\le\sum_{e\in E}w_ex_e$. Minimizing over feasible $x$ proves the lemma.
\end{proof}

\begin{theorem}
\label{thm:rank-reduction-gap}
Suppose $d\ge1$ satisfies $\lambda_w(N)\le d\,\sigma_w(N)$ for every positive-rank restriction $N$ of $M$.
Repeatedly deleting a minimum-weight cocircuit for $p$ steps produces a $p$-cocycle $S$ with $w(S)\le d\,\lambda_{p,w}^*(M)$,
where $\lambda_{p,w}^*(M)$ is the optimum of~\eqref{eq:base-hitting}.
\end{theorem}

\begin{proof}
Put $M_0:=M$. For $i=0,\ldots,p-1$, let $C_i$ be a minimum-weight cocircuit of $M_i$ and put $M_{i+1}:=M_i\setminus C_i$. The sets $E(M_i)$ form a chain of flats of $M$ with ranks $r(M)-i$, so $S:=\bigcup_{i=0}^{p-1}C_i$ is a $p$-cocycle. Since $\lambda_w(M_i)\le d\,\sigma_w(M_i)$, Lemma~\ref{lem:cocircuit-lp-drop} gives
\[
\begin{aligned}
w(S)&=\sum_{i=0}^{p-1}w(C_i)
\le d\sum_{i=0}^{p-1}\lambda_{1,w}^*(M_i)\\
&\le d\sum_{i=0}^{p-1}\bigl(\lambda_{p-i,w}^*(M_i)-\lambda_{p-i-1,w}^*(M_{i+1})\bigr)\\
&=d\,\lambda_{p,w}^*(M).
\end{aligned}
\]
\end{proof}

If every member $M$ of a deletion-closed class satisfies $\cogirthratiowt(M)\le d$, then the weighted LP gap on that class is at most $d$.

\subsection{Enumeration from the LP gap}

The next theorem adapts the LP enumeration of Chekuri, Quanrud, and Xu~\cite{chekuri_lp_2020}, with bases in place of spanning trees and flats in place of graph components.

\begin{theorem}
\label{thm:lp-enumeration}
Given an independence oracle for $M$, a minimum-weight $p$-cocycle is computable deterministically in $n^{(2\rho_{p,w}-1)p+O(1)}$ time,
where $\rho_{p,w}:=\frac{\lambda_{p,w}(M)}{\lambda_{p,w}^*(M)}$ is not part of the input.
\end{theorem}

\begin{proof}
Assume $\lambda_{p,w}^*(M)>0$.
First we compute $\lambda_{p,w}^*(M)$ and an optimal dual solution $(y,z)$ of~\eqref{eq:base-packing}.
Separation over the base constraints of~\eqref{eq:base-hitting} computes a minimum-$x$-weight base
and reports a violation exactly when its $x$-weight is below $p$,
i.e., there exists a base $B$ of $M$ with $x(B) < p$.
So Schrijver's optimization--separation theorem~\cite{schrijver_combinatorial_2003}
recovers an optimal dual supported on at most $n$ base constraints.
Since the dual LP has $n$ constraints,
we may choose an optimal basic feasible solution $(y,z)$ with at most $n$ positive variables.
Let $\B^+:=\{B\in\B(M):y_B>0\}$ and $\beta:=\sum_B y_B$. Then $|\B^+|\le n$, and by strong duality, $\lambda_{p,w}^*(M)=p\beta-\sum_e z_e\le p\beta$.
Since $\lambda_{p,w}^*(M)>0$, we obtain $\beta\ge \lambda_{p,w}^*(M)/p>0$.

Fix any $p$-cocycle $A$, so $w(A)\ge \lambda_{p,w}^*(M)$. Sampling $B\in\B^+$ with probability $y_B/\beta$, dual feasibility gives
\[
\mathbb E\bigl[|A\cap B|\bigr]
=\frac1\beta\sum_{e\in A}\sum_{B\ni e}y_B
\le\frac{w(A)+z(A)}\beta
\le p+\frac{w(A)-\lambda_{p,w}^*(M)}{\beta}
\le\frac{p\,w(A)}{\lambda_{p,w}^*(M)},
\]
so some $B\in\B^+$ has $|A\cap B|\le\lfloor p\,w(A)/\lambda_{p,w}^*(M)\rfloor$.

Every $p$-cocycle meets every base in at least $p$ elements.
Fix $B\in\B^+$ and $S\subseteq B$ with $|S|\ge p$. We list all $p$-cocycles $A$ with $A\cap B=S$ as follows.
For such an $A$, the flat $E(M)\setminus A$ contains $B\setminus S$,
so $F:=E(M)\setminus(A\cup B)$ is a flat of $M/(B\setminus S)$ disjoint from $S$ with
\[
r_{M/(B\setminus S)}(F)=r_M(E(M)\setminus A)-r_M(B\setminus S)\le(r(M)-p)-(r(M)-|S|)=|S|-p.
\]
Conversely, for every such flat $F$, the set $(B\setminus S)\cup F$ is a flat of $M$ of rank at most $r(M)-p$, and its complement is a $p$-cocycle $A$ with $A\cap B=S$.
Every flat of rank at most $|S|-p$ is the closure of an independent set of size at most $|S|-p$,
so computing the closures of all such sets in $M/(B\setminus S)$ lists all such $A$ in $n^{|S|-p+O(1)}$ time.

For $t=p,p+1,\ldots$,
list the $p$-cocycles $A$ with $|A\cap B|\le t$ for some $B\in\B^+$,
let $m_t$ be the minimum weight found so far, and halt once $t\ge\lfloor p\,m_t/\lambda_{p,w}^*(M)\rfloor$.
By the averaging bound, every $p$-cocycle $A$ is listed by threshold $\lfloor p\,w(A)/\lambda_{p,w}^*(M)\rfloor$.
Hence, when the algorithm halts, every $p$-cocycle of weight at most $m_t$ has been listed, so $m_t=\lambda_{p,w}(M)$.
Moreover, a minimum-weight $p$-cocycle is listed by threshold $\lfloor p\rho_{p,w}\rfloor$, after which $m_t=\lambda_{p,w}(M)$, so the algorithm halts by $t=\lfloor p\rho_{p,w}\rfloor$.
Threshold $t$ costs $n\cdot n^t\cdot n^{t-p+O(1)}$,
and summing up to $t=\lfloor p\rho_{p,w}\rfloor$ gives $n^{(2\rho_{p,w}-1)p+O(1)}$ time.
\end{proof}

Suppose $\lambda_w(N)\le d\,\sigma_w(N)$ for every positive-rank restriction $N$ of $M$.
Then $\rho_{p,w}\le d$ by Theorem~\ref{thm:rank-reduction-gap}, so Theorem~\ref{thm:lp-enumeration} solves rank-$p$ reduction exactly in $n^{(2d-1)p+O(1)}$ time.
Moreover, each step of the greedy algorithm in Theorem~\ref{thm:rank-reduction-gap} needs a minimum-weight cocircuit of some restriction $N$, which Theorem~\ref{thm:lp-enumeration} with $p=1$ finds in $n^{2d-1+O(1)}$ time since $\rho_{1,w}=\lambda_w(N)/\sigma_w(N)\le d$.
Hence the greedy $d$-approximation runs in $p\,n^{2d-1+O(1)}$ time. Neither algorithm needs to know $d$.

For unit weights, the case $p=1$ also decides in polynomial time whether $\lambda(M)$ equals the maximum number of disjoint bases of $M$, which is $\lfloor\sigma(M)\rfloor$ by Edmonds' base packing theorem~\cite{edmonds_lehmans_1965}; this resolves the recognition problem noted by Bailey, Newman, and Stevens~\cite{BaileyNS14}.
Since $\lambda(M)\ge\sigma(M)$, equality holds only if $\sigma(M)$ is an integer and $\lambda(M)=\sigma(M)$.
In that case every minimum cocircuit $A$ has $|A|=\sigma(M)=\lambda_{1}^*(M)$, so the averaging bound in the proof of Theorem~\ref{thm:lp-enumeration} shows that $A$ meets some $B\in\B^+$ in exactly one element.
Hence it suffices to run the enumeration with threshold $t=1$, which takes polynomial time, and compare the minimum found with $\sigma(M)$.

\section{Matroid union}
\label{sec:matroid-union}

For matroids $M_1,\ldots,M_m$ on a common ground set $E$,
their matroid union $M_1\vee\cdots\vee M_m$ is a matroid on $E$ with independent sets
$\{I_1\cup\cdots\cup I_m : I_i\in\mathcal I(M_i),\ 1\le i\le m\}$.

\begin{theorem}
\label{thm:ratio-subadditivity-under-union}
Let $M_1,\ldots,M_m$ be matroids on the same ground set $E$, and put $M:=M_1\vee\cdots\vee M_m$. Then
\[
\cogirthratio(M)\le\sum_{i=1}^m\cogirthratio(M_i).
\]
\end{theorem}

\begin{proof}
Put $h:=\sum_i1/\sigma(M_i)$. We first prove $\sigma(M)\ge1/h$.
If $h\ge1$, this follows from $\sigma(M)\ge1$. Assume $h<1$,
and write $r,r_i$ for the rank functions of $M,M_i$.
For $A\subseteq E$, choose $Y\subseteq A$ attaining the
matroid-union rank formula $r(A)=|A\setminus Y|+\sum_i r_i(Y)$~\cite[Theorem~12.3.1]{Oxley06}.
The strength formula gives $r_i(E)-r_i(Y)\le |E\setminus Y|/\sigma(M_i)$, also when the left side is zero.
Since $r(E)\le\sum_i r_i(E)$,
\[
\begin{aligned}
r(E)-r(A)
&\le\sum_i\bigl(r_i(E)-r_i(Y)\bigr)-|A\setminus Y|\\
&\le h|E\setminus Y|-|A\setminus Y|\\
&=h|E\setminus A|+(h-1)|A\setminus Y|\\
&\le h|E\setminus A|.
\end{aligned}
\]
Taking $A$ to be a flat of $M$ with $r(A)<r(E)$ gives $|E\setminus A|/(r(E)-r(A))\ge1/h$, so $\sigma(M)\ge1/h$.

Every base of $M$ spans each $M_i$: otherwise, an element outside its $M_i$-closure could be added to its $M_i$-part in a union decomposition. Thus every cocircuit of $M_i$ meets every base of $M$ and hence contains a cocircuit of $M$, giving $\lambda(M)\le\lambda(M_i)$. Combining the two bounds yields $\cogirthratio(M)\le\lambda(M)\sum_i1/\sigma(M_i)\le\sum_i\cogirthratio(M_i)$.
\end{proof}

Write $D(M):=\max_{X\subseteq E(M)} \{\cogirthratio(M|X):r_M(X)>0\}$. Restriction commutes with matroid union, so applying the theorem to the positive-rank restrictions of the summands gives $D(M_1\vee\cdots\vee M_m)\le\sum_iD(M_i)$. Thus the greedy algorithm of Theorem~\ref{thm:rank-reduction-gap} gives a $\sum_iD(M_i)$-approximation for unweighted rank reduction on the union.

\section{Lower-range count matroids}

Throughout this section, fix integers $k\ge1$, $0\le \ell\le k$, and $s\ge2$, and let $H=(V,E)$ be a hypergraph.
For $I\subseteq E$, let $V(I)$ be the set of vertices met by $I$.
Recall that the $(k,\ell)$-count matroid $M_{k,\ell}(H)$ has as its independent sets
those $I\subseteq E$ satisfying $|J|\le k|V(J)|-\ell$ for every nonempty $J\subseteq I$.
We write $K_v^{(s)}$ for the complete $s$-uniform hypergraph on $v$ vertices, and $K_v:=K_v^{(2)}$.

\begin{theorem}
\label{thm:count-unit-bound}
If $H$ has edge size at most $s$, then $\cogirthratio(M_{k,\ell}(H))\le ks$. Moreover, $\cogirthratio(M_{k,\ell}(K_v^{(s)}))\to ks$ as $v\to\infty$.
\end{theorem}

Since lower-range count matroids are closed under deletion,
Theorem~\ref{thm:rank-reduction-gap} and Theorem~\ref{thm:lp-enumeration},
with unit weights, give a $ks$-approximation for every feasible rank loss $p\ge1$
and exact rank-$p$ reduction in $n^{(2ks-1)p+O(1)}$ time.

By the maps-and-trees decomposition~\cite[Theorem~1.7 and Proposition~4.3]{streinu_sparse_hypergraphs_2009}, $M_{k,\ell}(H)$ is the matroid union of $\ell$ copies of the hypergraphic matroid $M_{1,1}(H)$ and $k-\ell$ copies of $M_{1,0}(H)$, the transversal matroid of the edge--vertex incidence bipartite graph:
\begin{equation}\label{eq:count-union}
M_{k,\ell}(H)=\bigvee^{\ell}M_{1,1}(H)\vee\bigvee^{k-\ell}M_{1,0}(H).
\end{equation}
By Theorem~\ref{thm:ratio-subadditivity-under-union}, the upper bound $\cogirthratio(M_{k,\ell}(H))\le ks$ reduces to bounding the ratios of $M_{1,1}(H)$ and $M_{1,0}(H)$ by $s$.

\subsection{The upper bound}

For $M_{1,1}(H)$, we translate the known hypergraph partition bound into a cogirth--strength bound.

\begin{proposition}
\label{prop:hypergraphic-unit-atom}
If $H$ has edge size at most $s$, then $\cogirthratio(M_{1,1}(H))\le s$.
\end{proposition}

\begin{proof}
Write $M:=M_{1,1}(H)$, $r:=r_M$, and $v:=|V(H)|$. Since $r(M)>0$, we have $v\ge2$.
For a partition $\mathcal P$ of $V(H)$,
let $\delta_H(\mathcal P)$ be the set of edges meeting more than one part,
and set $\tau(H):=\min_{|\mathcal P|\ge2}|\delta_H(\mathcal P)|/(|\mathcal P|-1)$.
By the hypergraphic rank formula of Frank, Király, and Kriesell~\cite{frank_decomposing_2003},
$r(A)=\min_{\mathcal P}\bigl(v-|\mathcal P|+|A\cap\delta_H(\mathcal P)|\bigr)$ for every $A\subseteq E$.
In particular, $r(E)\le v-1$, with equality if and only if $\tau(H)\ge1$.

Suppose first that $\tau(H)\ge1$, so $r(E)=v-1$. We show that $\sigma(M)=\tau(H)$.
For any $A\subseteq E$ with $r(A)<r(E)$,
choose a partition $\mathcal P$ attaining $r(A)=v-|\mathcal P|+|A\cap\delta_H(\mathcal P)|$.
Since $|\delta_H(\mathcal P)|\ge|\mathcal P|-1$, subtracting $|A\cap\delta_H(\mathcal P)|$ from numerator and denominator shows
\[
\frac{|E\setminus A|}{r(E)-r(A)}\ge\frac{|\delta_H(\mathcal P)\setminus A|}{|\mathcal P|-1-|A\cap\delta_H(\mathcal P)|}\ge\frac{|\delta_H(\mathcal P)|}{|\mathcal P|-1}\ge\tau(H).
\]
Conversely, let $\mathcal P$ be a partition attaining $\tau(H)$, and put $A:=E\setminus\delta_H(\mathcal P)$.
Since $r(A)\le v-|\mathcal P|$, we have $r(E)-r(A)\ge|\mathcal P|-1$, and
taking $F:=\operatorname{cl}_M(A)$ gives
\[
\sigma(M) \le\frac{|E\setminus F|}{r(E)-r(F)} \le\frac{|\delta_H(\mathcal P)|}{|\mathcal P|-1} =\tau(H).
\]
Hence $\sigma(M)=\tau(H)$.
By a result of Baïou and Barahona~\cite{baiou_packing_2022}, there is a bipartition $\mathcal P$ of $V(H)$ with $|\delta_H(\mathcal P)|\le s(1-1/v)\tau(H)$.
Deleting $\delta_H(\mathcal P)$ lowers the rank of $M$ to at most $v-2$, so $\lambda(M)\le|\delta_H(\mathcal P)|<s\sigma(M)$.

Having established the bound when $\tau(H)\ge1$, we now consider general $H$.
Choose a partition $\mathcal P$ attaining $r(E)=v-|\mathcal P|+|\delta_H(\mathcal P)|$,
and for each $X\in\mathcal P$ let $H[X]:=(X,E[X])$ be the subhypergraph of edges contained in $X$.
Since $r(E[X])\le|X|-1$ for each $X\in\mathcal P$, the sets $E[X]$ contribute at most $\sum_{X\in\mathcal P}(|X|-1)=v-|\mathcal P|$ to the rank, and each edge in $\delta_H(\mathcal P)$ contributes at most one.
Equality implies that $r(E[X])=|X|-1$ for every $X\in\mathcal P$, every edge in $\delta_H(\mathcal P)$ is a coloop,
and $M$ is the direct sum of these coloops and the restrictions $M|E[X]=M_{1,1}(H[X])$ with $E[X]\ne\emptyset$.
Each such $H[X]$ has $|X|\ge2$ and hence $\tau(H[X])\ge1$, so $\cogirthratio(M_{1,1}(H[X]))\le s$ by the first case,
while each coloop has cogirth--strength ratio $1$.
Hence, Lemma~\ref{lem:direct-sum} gives $\cogirthratio(M)\le s$.
\end{proof}

For $M_{1,0}$, the bound follows from the transversal-matroid rank formula.

\begin{proposition}
\label{prop:incidence-transversal-unit-atom}
If $H$ has edge size at most $s$, then $\cogirthratio(M_{1,0}(H))\le s$.
\end{proposition}

\begin{proof}
Write $M:=M_{1,0}(H)$ and $V:=V(E)$.
The defect form of Hall's theorem gives the standard transversal-matroid rank formula
$r_M(A)=\min_{Y\subseteq A}(|A\setminus Y|+|V(Y)|)$~\cite[Theorem~22.2]{schrijver_combinatorial_2003}.
Let $Y_0$ attain the minimum for $A=E$.
Then $r_M(Y_0)=|V(Y_0)|$, otherwise some subset $Y'\subseteq Y_0$ would yield a smaller value in the rank formula for $E$.
Hence $r(M)=r_M(Y_0)+|E\setminus Y_0|$, so every element of $E\setminus Y_0$ is a coloop.
If such an element exists, then $\lambda(M)=\sigma(M)=1$. We may therefore assume $Y_0=E$ and $r(M)=|V|=:v$.

Take $A\subseteq E$ with $r_M(A)<v$, and let $Y\subseteq A$ attain the minimum in the rank formula for $A$.
Put $U:=V\setminus V(Y)$. Then $v-r_M(A)=|U|-|A\setminus Y|>0$.
Applying the rank formula for $E$ to $Y$ gives $|E\setminus A|\ge v-r_M(A)$.
Deleting all edges incident to any vertex in $U$ reduces the rank,
so $\lambda(M)$ is at most the minimum number of edges incident to a vertex in $U$.
Since no edge of $Y$ meets $U$, and each edge of $E\setminus Y$ meets at most $s$ vertices of $U$, averaging gives
\[
\lambda(M)\le \frac{s|E\setminus Y|}{|U|}=\frac{s(|E\setminus A|+|A\setminus Y|)}{v-r_M(A)+|A\setminus Y|}\le\frac{s|E\setminus A|}{v-r_M(A)},
\]
where the last inequality follows from $|E\setminus A|\ge v-r_M(A)$. Minimizing over such $A$ proves the proposition.
\end{proof}

Propositions~\ref{prop:hypergraphic-unit-atom} and~\ref{prop:incidence-transversal-unit-atom}, together with \eqref{eq:count-union} and Theorem~\ref{thm:ratio-subadditivity-under-union}, prove the upper bound $\cogirthratio(M_{k,\ell}(H))\le ks$ in Theorem~\ref{thm:count-unit-bound}. We defer the matching lower bound to Appendix~\ref{app:sharpness}: the complete $s$-uniform hypergraphs $K_v^{(s)}$ satisfy $\cogirthratio(M_{k,\ell}(K_v^{(s)}))\to ks$ as $v\to\infty$. Adding $p-1$ vertex-disjoint single-edge components to $K_v^{(s)}$, which contribute $p-1$ coloops, yields count matroids of hypergraphs with edge size at most $s$ whose rank-$p$ integrality gap $\lambda_p/\lambda_p^*$ also tends to $ks$. Hence the LP gap bound $ks$ is sharp for every fixed $p$.

Although our unweighted analysis does not generalize directly to the weighted case, it is natural to ask whether a bound independent of the weights still holds for all lower-range count matroids of bounded edge size.
However, no such bound exists.
In fact, the weighted cogirth--strength ratio, and hence the weighted integrality gap of~\eqref{eq:base-hitting}, is unbounded even for $(2,2)$-count matroids of simple graphs, whose bases are disjoint unions of two spanning trees.

\begin{figure}[!htbp]
\centering
\includegraphics[width=0.85\textwidth]{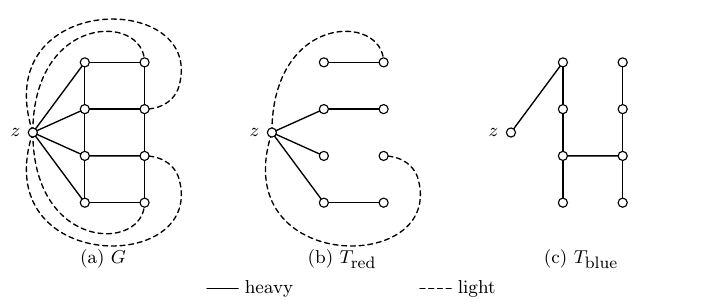}
\caption{The construction for $q=2$, with columns $C_1,C_2$ from left to right.
Selecting the light edges in the first and third rows gives two edge-disjoint spanning trees $T_{\mathrm{red}}$ and $T_{\mathrm{blue}}$, whose union consists of all heavy edges and the two selected light edges.}
\label{fig:weighted-count-example}
\end{figure}

For any integer $q\ge2$, form a simple graph $G$ from a grid with $2q$ rows and $q$ columns $C_1,\ldots,C_q$ by adding a vertex $z$ adjacent to both ends of every row.
Give the $2q$ edges between $z$ and $C_q$ weight~$1$ and call them \emph{light}; give all other edges weight~$q+1$ and call them \emph{heavy}.
We claim that the heavy edges together with any $q$ light edges decompose into two edge-disjoint spanning trees.
Indeed, color the horizontal edges and the $2q$ edges between $z$ and $C_1$ red, and color the vertical edges blue; this partitions the heavy edges into a red spanning tree of size $2q^2$ and a blue forest of size $2q^2-q$ with components $z,C_1,\ldots,C_q$.
Adding any $q$ light edges to the red tree creates $q$ edge-disjoint cycles, each formed by $z$ and one chosen row.
Assign these $q$ rows bijectively to $C_1,\ldots,C_q$, and in the row assigned to $C_j$, recolor blue the red edge entering $C_j$ from $C_{j-1}$ (or from $z$ if $j=1$).
This breaks each red cycle while linking the blue components along $z,C_1,\ldots,C_q$ into a spanning tree, proving the claim.
Figure~\ref{fig:weighted-count-example} illustrates the construction and the decomposition for $q=2$.

Thus the heavy edges plus any $q$ light edges form a base of $M:=M_{2,2}(G)$, so $r(M)=4q^2$, and since there are $4q^2-q$ heavy edges, every base uses at least $q$ light edges.
Hence $\lambda_w(M)=q+1$.
Meanwhile, since every base uses at least $q$ of the $2q$ light edges, $\sigma_w(M)\le 2$, and splitting the light edges into two sets of size $q$ gives two bases within capacity $w$, so $\sigma_w(M)=2$ and $\cogirthratiowt_w(M)=(q+1)/2$.

\section{Quasi-graphic matroids}
\label{sec:quasigraphic}

Quasi-graphic matroids include all frame and lifted-graphic matroids~\cite{bowler_describing_2025}. In particular, for a graph $G$, both the graphic matroid $M_{1,1}(G)$ and the bicircular matroid $M_{1,0}(G)$ are quasi-graphic. We use the tripartition description of Bowler, Funk, and Slilaty~\cite{bowler_describing_2025}.

Let $G$ be a graph. A theta subgraph of $G$ consists of two distinct vertices together with three internally disjoint paths between them. A \emph{proper tripartition} of the cycles of $G$ is a partition $(\mathcal B,\mathcal L,\mathcal F)$ of its cycles into \emph{balanced}, \emph{lift-type}, and \emph{frame-type} cycles such that
\begin{itemize}
\item no theta subgraph has exactly two of its three cycles in $\mathcal B$, and
\item every cycle in $\mathcal L$ shares a vertex with every cycle in $\mathcal F$.
\end{itemize}
A subgraph is balanced if all its cycles lie in $\mathcal B$; in particular, an acyclic subgraph is balanced. For $S\subseteq E(G)$, consider the spanning subgraph $(V(G),S)$. Let $c(S)$ be its number of components, let $b(S)$ be its number of balanced components, counting isolated vertices, and let $\eta(S)=1$ if $S$ contains a cycle in $\mathcal L$ and $\eta(S)=0$ otherwise.

The associated quasi-graphic matroid has rank function~\cite[Lemma~2.6]{bowler_describing_2025}
\[
r(S)=
\begin{cases}
|V(G)|-b(S), & \text{if }S\text{ contains a cycle in }\mathcal F,\\
|V(G)|-c(S)+\eta(S), & \text{otherwise}.
\end{cases}
\]

\begin{theorem}
\label{thm:quasi-graphic-ratio}
Every quasi-graphic matroid $M$ satisfies $\cogirthratiowt(M)\le 2$.
\end{theorem}

\begin{proof}
Fix weights $w:E(M)\to\mathbb R_{\ge 0}$ with $\sigma_w(M)>0$. By Lemma~\ref{lem:direct-sum} we may assume that $M$ is connected. If $M$ is graphic, Nash--Williams--Tutte~\cite{Nash-Williams_1961,Tutte_1961} gives the bound directly. Otherwise, the connected-framework reduction of Bowler, Funk, and Slilaty~\cite{bowler_describing_2025} gives a connected graph $G=(V,E)$ and a proper tripartition representing $M$. Its framework contains an unbalanced cycle, so $r(E)=|V|$. Put $r:=r(E)$ and $\lambda_w:=\lambda_w(M)$.

We prove that $2w(E\setminus S)\ge h\lambda_w$ for every $S\subseteq E$ with $h:=r-r(S)>0$. Taking the minimum of ${w(E\setminus S)}/{h}$ over $S$ then gives $\sigma_w(M)\ge\lambda_w(M)/2$.

For $X\subseteq V$, let $E[X]$ be the edges with both endpoints in $X$, let $S[X]:=S\cap E[X]$, and let $\delta(X)$ be the edges with exactly one endpoint in $X$. Let $V_1,\ldots,V_q$ be the vertex sets of the components of $(V,S)$. For each component set $C_i:=(E[V_i]\setminus S[V_i])\cup\delta(V_i)\subseteq E\setminus S$. Each edge of $E\setminus S$ lies in at most two of these sets. An edge within one component, including a loop, lies in one $C_i$, while an edge joining two components lies in two.

Deleting $C_i$ leaves $V_i$ as a separated component carrying exactly $S[V_i]$. If this component is balanced, the rank drops: a surviving frame-type cycle makes the first line of the rank formula at most $|V|-1$, and otherwise the second line gives the same bound. In the latter case there are at least two components when $V_i\subsetneq V$, while $V_i=V$ leaves a connected balanced graph of rank $|V|-1$.

Suppose $S$ contains no frame-type cycle and $S[V_i]$ is unbalanced. It contains a lift-type cycle, which meets every frame-type cycle. Deleting $C_i$ therefore destroys all frame-type cycles: none remains inside $V_i$, outside it, or across its boundary. Moreover, $V_i\subsetneq V$, since otherwise $S$ would have full rank. The surviving graph has at least two components, so the second line of the rank formula gives rank at most $|V|-1$.

If $S$ contains a frame-type cycle, charge its balanced components, whose number is $h=b(S)$. Otherwise charge all $q$ components, where $h=q-\eta(S)\le q$. Every charged $C_i$ contains a cocircuit.

In all cases, at least $h$ components are charged, and each charged $C_i$ has weight at least $\lambda_w$. The multiplicity bound gives
\[
h\lambda_w\le\sum_{i\text{ charged}}w(C_i)\le2w(E\setminus S),
\]
as required.
\end{proof}

The constant $2$ is asymptotically sharp already for graphic matroids: the graphic matroid $M$ of $K_v$ has $\lambda(M)=v-1$ and $\sigma(M)=v/2$, so $\cogirthratio(M)=2-2/v\to2$.

The quasi-graphic rank formula remains valid after deleting edges, so quasi-graphic matroids form a deletion-closed class. With Theorem~\ref{thm:quasi-graphic-ratio}, Theorem~\ref{thm:rank-reduction-gap} and Theorem~\ref{thm:lp-enumeration} give a $2$-approximation and an exact $n^{3p+O(1)}$-time algorithm for weighted rank-$p$ reduction.

\paragraph*{Acknowledgements}
Supported by the National Natural Science Foundation of China under
grant 62372093 and by the CCF-Huawei Populus Grove Fund -- Theoretical
Computer Science Special Project CCF-HuaweiLK2025003.

We thank Benjamin Moseley for hosting Chao at the CMU Tepper School
of Business, where part of this work was carried out.

\paragraph*{AI disclosure}
The quasi-graphic result in Section~\ref{sec:quasigraphic} was discovered by Fable 5, and the manuscript was prepared with assistance from AI tools.
The authors take full responsibility for
the correctness and content of the paper, including material generated
with AI assistance.

\bibliographystyle{plain}
\bibliography{ref}

@article{geelen_computing_2018,
  author        = {Geelen, Jim and Kapadia, Rohan},
  title         = {Computing {Girth} and {Cogirth} in {Perturbed} {Graphic} {Matroids}},
  journal       = {Combinatorica},
  volume        = {38},
  number        = {1},
  pages         = {167--191},
  year          = {2018},
  month         = {feb},
  issn          = {0209-9683, 1439-6912},
  doi           = {10.1007/s00493-016-3445-3},
  url           = {https://link.springer.com/10.1007/s00493-016-3445-3}
}

@article{streinu_sparse_hypergraphs_2009,
  author        = {Streinu, Ileana and Theran, Louis},
  title         = {Sparse Hypergraphs and Pebble Game Algorithms},
  journal       = {European Journal of Combinatorics},
  volume        = {30},
  number        = {8},
  pages         = {1944--1964},
  year          = {2009},
  doi           = {10.1016/j.ejc.2008.12.018}
}

@article{BaileyNS14,
  author        = {Bailey, Robert F. and Newman, Mike and Stevens, Brett},
  title         = {A note on packing spanning trees in graphs and bases in matroids},
  journal       = {Australas. J. Combin.},
  volume        = {59},
  pages         = {24--38},
  year          = {2014},
  issn          = {1034-4942},
  url           = {https://ajc.maths.uq.edu.au/pdf/59/ajc_v59_p024.pdf}
}

@book{Oxley06,
  author        = {Oxley, James G.},
  title         = {Matroid Theory},
  series        = {Oxford graduate texts in mathematics},
  year          = {2006},
  publisher     = {Oxford University Press},
  isbn          = {9780199202508}
}

@techreport{Kiraly09,
  author        = {Kir{\'a}ly, Tam{\'a}s},
  title         = {Computing the minimum cut in hypergraphic matroids},
  number        = {QP-2009-05},
  year          = {2009},
  institution   = {Egerv{\'a}ry Research Group, Budapest},
  note          = {{\tt www.cs.elte.hu/egres}}
}

@article{JoretV15,
  author        = {Joret, Gwena{\"e}l and Vetta, Adrian},
  title         = {Reducing the rank of a matroid},
  journal       = {Discrete Mathematics \& Theoretical Computer Science},
  volume        = {17},
  number        = {2},
  pages         = {143--156},
  year          = {2015},
  month         = {sep},
  doi           = {10.46298/dmtcs.2135},
  url           = {https://dmtcs.episciences.org/2135}
}

@article{Karger98,
  author        = {Karger, David R},
  title         = {Random sampling and greedy sparsification for matroid optimization problems},
  journal       = {Mathematical Programming},
  volume        = {82},
  number        = {1-2},
  pages         = {41--81},
  year          = {1998},
  issn          = {0025-5610},
  doi           = {10.1007/BF01585865},
  url           = {https://people.csail.mit.edu/karger/Papers/matroid.pdf}
}

@book{schrijver_combinatorial_2003,
  author        = {Schrijver, Alexander},
  title         = {Combinatorial optimization: polyhedra and efficiency},
  series        = {Algorithms and combinatorics},
  number        = {24},
  year          = {2003},
  publisher     = {Springer},
  address       = {Berlin Heidelberg},
  isbn          = {978-3-540-44389-6}
}

@article{saran_finding_1995,
  author        = {Saran, Huzur and Vazirani, Vijay V.},
  title         = {Finding {$k$} Cuts within Twice the Optimal},
  journal       = {SIAM Journal on Computing},
  volume        = {24},
  number        = {1},
  pages         = {101--108},
  year          = {1995},
  doi           = {10.1137/S0097539792251730}
}

@article{chekuri_lp_2020,
  author        = {Chekuri, Chandra and Quanrud, Kent and Xu, Chao},
  title         = {{LP} {Relaxation} and {Tree} {Packing} for {Minimum} {$k$}-{Cut}},
  journal       = {SIAM Journal on Discrete Mathematics},
  volume        = {34},
  number        = {2},
  pages         = {1334--1353},
  year          = {2020},
  issn          = {0895-4801, 1095-7146},
  doi           = {10.1137/19M1299359}
}

@book{frank_connections_2011,
  author        = {Frank, Andr{\'a}s},
  title         = {Connections in {Combinatorial} {Optimization}},
  year          = {2011},
  publisher     = {Oxford University Press},
  isbn          = {978-0-19-920527-1}
}

@article{catlin_fractional_1992,
  author        = {Catlin, Paul A. and Grossman, Jerrold W. and Hobbs, Arthur M. and Lai, Hong-Jian},
  title         = {Fractional arboricity, strength, and principal partitions in graphs and matroids},
  journal       = {Discrete Applied Mathematics},
  volume        = {40},
  number        = {3},
  pages         = {285--302},
  year          = {1992},
  issn          = {0166218X},
  doi           = {10.1016/0166-218X(92)90002-R}
}

@article{Nash-Williams_1961,
  author        = {Nash-Williams, C. St. J. A.},
  title         = {Edge-Disjoint Spanning Trees of Finite Graphs},
  journal       = {Journal of the London Mathematical Society},
  volume        = {s1-36},
  number        = {1},
  pages         = {445--450},
  year          = {1961},
  issn          = {00246107},
  doi           = {10.1112/jlms/s1-36.1.445}
}

@article{Tutte_1961,
  author        = {Tutte, W. T.},
  title         = {On the Problem of Decomposing a Graph into $n$ Connected Factors},
  journal       = {Journal of the London Mathematical Society},
  volume        = {s1-36},
  number        = {1},
  pages         = {221--230},
  year          = {1961},
  doi           = {10.1112/jlms/s1-36.1.221}
}

@misc{bowler_describing_2025,
  author        = {Bowler, Nathan and Funk, Daryl and Slilaty, Daniel},
  title         = {Describing Quasi-Graphic Matroids},
  year          = {2025},
  eprint        = {1808.00489},
  archiveprefix = {arXiv},
  primaryclass  = {math.CO},
  url           = {https://arxiv.org/abs/1808.00489v4},
  note          = {Version 4, incorporating corrections}
}

@article{goldschmidt_polynomial_1994,
  author        = {Goldschmidt, Olivier and Hochbaum, Dorit S.},
  title         = {A Polynomial Algorithm for the $k$-Cut Problem for Fixed $k$},
  journal       = {Mathematics of Operations Research},
  volume        = {19},
  number        = {1},
  pages         = {24--37},
  year          = {1994},
  doi           = {10.1287/moor.19.1.24}
}

@inproceedings{thorup_minimum_2008,
  author        = {Thorup, Mikkel},
  title         = {Minimum $k$-way cuts via deterministic greedy tree packing},
  booktitle     = {Proceedings of the 40th Annual ACM Symposium on Theory of Computing},
  pages         = {159--166},
  year          = {2008},
  doi           = {10.1145/1374376.1374402}
}

@article{gupta_optimal_2022,
  author        = {Gupta, Anupam and Harris, David G. and Lee, Euiwoong and Li, Jason},
  title         = {Optimal Bounds for the $k$-cut Problem},
  journal       = {Journal of the ACM},
  volume        = {69},
  number        = {1},
  pages         = {1--18},
  year          = {2022},
  doi           = {10.1145/3478018}
}

@inproceedings{GuptaLL18,
  author        = {Gupta, Anupam and Lee, Euiwoong and Li, Jason},
  title         = {Faster exact and approximate algorithms for $k$-cut},
  booktitle     = {59th IEEE Annual Symposium on Foundations of Computer Science (FOCS)},
  pages         = {113--123},
  year          = {2018},
  doi           = {10.1109/FOCS.2018.00020}
}

@inproceedings{karger_global_1993,
  author        = {Karger, David R.},
  title         = {Global min-cuts in {RNC}, and other ramifications of a simple min-cut algorithm},
  booktitle     = {Proceedings of the 4th Annual {ACM}-{SIAM} Symposium on Discrete Algorithms},
  pages         = {21--30},
  year          = {1993}
}

@article{frank_decomposing_2003,
  author        = {Frank, Andr{\'a}s and Kir{\'a}ly, Tam{\'a}s and Kriesell, Matthias},
  title         = {On decomposing a hypergraph into $k$ connected sub-hypergraphs},
  journal       = {Discrete Applied Mathematics},
  volume        = {131},
  number        = {2},
  pages         = {373--383},
  year          = {2003},
  doi           = {10.1016/S0166-218X(02)00463-8}
}

@article{narayanan_approximation_1996,
  author        = {Narayanan, H. and Roy, Subir K. and Patkar, Sachin B.},
  title         = {Approximation Algorithms for Min-$k$-Overlap Problems Using the Principal Lattice of Partitions Approach},
  journal       = {Journal of Algorithms},
  volume        = {21},
  number        = {2},
  pages         = {306--330},
  year          = {1996},
  doi           = {10.1006/jagm.1996.0047}
}

@article{fukunaga_computing_2013,
  author        = {Fukunaga, Takuro},
  title         = {Computing Minimum Multiway Cuts in Hypergraphs},
  journal       = {Discrete Optimization},
  volume        = {10},
  number        = {4},
  pages         = {371--382},
  year          = {2013},
  doi           = {10.1016/j.disopt.2013.10.002}
}

@inproceedings{baiou_packing_2022,
  author        = {Ba{\"i}ou, Mourad and Barahona, Francisco},
  title         = {Packing Hypertrees and the $k$-Cut Problem in Hypergraphs},
  booktitle     = {Learning and Intelligent Optimization},
  series        = {Lecture Notes in Computer Science},
  volume        = {13621},
  pages         = {521--534},
  publisher     = {Springer International Publishing},
  address       = {Cham},
  year          = {2022},
  isbn          = {978-3-031-24866-5},
  doi           = {10.1007/978-3-031-24866-5_37}
}

@inproceedings{Li_2019,
  title         = {Faster Minimum k-cut of a Simple Graph},
  booktitle     = {2019 IEEE 60th Annual Symposium on Foundations of Computer Science (FOCS)},
  publisher     = {IEEE},
  author        = {Li, Jason},
  year          = {2019},
  pages         = {1056--1077},
  doi           = {10.1109/FOCS.2019.00068}
}

@misc{cong_derandomizing_2026,
  author        = {Cong, Yu and Xu, Chao and Zhao, Yajie},
  title         = {Derandomizing {Karger's} Contraction Algorithm for Matroids},
  year          = {2026},
  eprint        = {2608.16298},
  archiveprefix = {arXiv},
  howpublished  = {arXiv:2608.16298}
}

@article{karger_new_1996,
  author        = {Karger, David R. and Stein, Clifford},
  title         = {A new approach to the minimum cut problem},
  journal       = {Journal of the ACM},
  volume        = {43},
  number        = {4},
  pages         = {601--640},
  year          = {1996},
  doi           = {10.1145/234533.234534}
}

@misc{chandrasekaran_random_2026,
  author        = {Chandrasekaran, Karthekeyan and Chekuri, Chandra and Zhu, Weihao},
  title         = {Random Deletion for Circuits in Regular Matroids},
  year          = {2026},
  eprint        = {2610.33918},
  archiveprefix = {arXiv},
  howpublished  = {arXiv:2610.33918}
}

@misc{cong_destroying_2026,
  author        = {Cong, Yu and Xu, Chao and Zhao, Yajie},
  title         = {Destroying Rigidity in the Plane},
  year          = {2026},
  eprint        = {2610.38396},
  archiveprefix = {arXiv},
  howpublished  = {arXiv:2610.38396}
}

@article{edmonds_lehmans_1965,
  author        = {Edmonds, Jack},
  title         = {Lehman's switching game and a theorem of {Tutte} and {Nash-Williams}},
  journal       = {Journal of Research of the National Bureau of Standards, Section B: Mathematics and Mathematical Physics},
  volume        = {69B},
  number        = {1--2},
  pages         = {73--77},
  year          = {1965}
}

\appendix

\section{Sharpness for lower-range count matroids}
\label{app:sharpness}

Throughout this appendix, fix integers $k\ge1$ and $0\le\ell\le k$. For a hypergraph $H=(V,E)$, $A\subseteq E$, and $X\subseteq V$, write $A[X]:=\{e\in A:e\subseteq X\}$. A \emph{vertex subpartition} of $H$ is a collection $\mathcal X=(X_1,\ldots,X_q)$ of pairwise disjoint nonempty subsets of $V$, including the empty collection ($q=0$).

\begin{lemma}
\label{lem:count-vertex-subpartition}
Let $H=(V,E)$ be a hypergraph such that $M:=M_{k,\ell}(H)$ is loopless. For every $A\subseteq E$,
\[
r_M(A)=\min_{\mathcal X=(X_1,\ldots,X_q)}\biggl(\Bigl|A\setminus\bigcup_{i=1}^q A[X_i]\Bigr|
+\sum_{i=1}^q(k|X_i|-\ell)\biggr),
\]
where $\mathcal X$ ranges over all vertex subpartitions of $H$.
\end{lemma}

\begin{proof}
By the count-matroid rank formula~\cite[Section~13.5.2]{frank_connections_2011},
\[
r_M(A)=\min_{(Y_1,\ldots,Y_p)}\biggl(\Bigl|A\setminus\bigcup_{j=1}^p Y_j\Bigr|
+\sum_{j=1}^p\bigl(k|V(Y_j)|-\ell\bigr)\biggr),
\]
where $(Y_1,\ldots,Y_p)$
ranges over collections of pairwise disjoint nonempty subsets of $A$,
including the empty collection. Starting from a minimizing collection $(Y_1,\ldots,Y_p)$,
merging any two parts with $V(Y_i)\cap V(Y_j)\ne\emptyset$ into $Y_i\cup Y_j$ leaves $A\setminus\bigcup_t Y_t$
unchanged and changes $\sum_t(k|V(Y_t)|-\ell)$ by $\ell-k|V(Y_i)\cap V(Y_j)|\le\ell-k\le0$.
Repeating until the supports $X_i:=V(Y_i)$ are pairwise disjoint produces a vertex subpartition
$\mathcal X=(X_1,\ldots,X_q)$. Each $X_i$ is nonempty since $M$ is loopless. Since $Y_i\subseteq A[X_i]$
for each $i$, we have $\bigl|A\setminus\bigcup_i A[X_i]\bigr|\le\bigl|A\setminus\bigcup_i Y_i\bigr|$,
so the value of $\mathcal X$ is at most $r_M(A)$.

Conversely, let $I\subseteq A$ be independent in $M$
and let $\mathcal X=(X_1,\ldots,X_q)$ be a vertex subpartition.
Since $M$ is loopless, every edge of $I$ is nonempty, so $I[X_1],\ldots,I[X_q]$ are pairwise disjoint.
Moreover, $|I[X_i]|\le k|X_i|-\ell$ for every $i$:
this holds by independence of $I$ when $I[X_i]\ne\emptyset$,
and because $k|X_i|-\ell\ge k-\ell\ge0$ when $I[X_i]=\emptyset$. Therefore
\[
|I|=\Bigl|I\setminus\bigcup_{i=1}^q I[X_i]\Bigr|+\sum_{i=1}^q|I[X_i]|
\le\Bigl|A\setminus\bigcup_{i=1}^q A[X_i]\Bigr|+\sum_{i=1}^q(k|X_i|-\ell),
\]
and taking the maximum over $I$ gives the reverse inequality.
\end{proof}

\begin{proposition}
\label{prop:complete-uniform-count-ratio}
For $s\ge3$, $\cogirthratio(M_{k,\ell}(K_v^{(s)}))\to ks$ as $v\to\infty$.
\end{proposition}

\begin{proof}
Write $M:=M_{k,\ell}(K_v^{(s)})$, and put $m:=\binom vs$, $R:=kv-\ell$,
$\Delta:=\binom{v-1}{s-1}$, and $\mu:=\Delta-kv$. We prove, for all sufficiently large $v$, that
$r(M)=R$, $\Delta-kv<\lambda(M)\le\Delta$, and $\sigma(M)=m/R$.

For a nonempty vertex subpartition \(\mathcal X=(X_1,\ldots,X_q)\) other than \(\{V\}\),
put \(u:=v-\sum_{i=1}^q|X_i|\) and \(c:=m-\sum_{i=1}^q\binom{|X_i|}{s}\).
Since the sets \(X_i\) are pairwise disjoint and nonempty,
and \(\mathcal X\neq\{V\}\),
discrete convexity of \(t\mapsto\binom{t}{s}\) yields \(\sum_{i=1}^q\binom{|X_i|}{s}\leq\binom{v-1}{s}\).
Indeed, the sum is maximized by concentrating the vertices into as few parts as possible,
with at least two nonempty parts.
Consequently, \(c\geq\binom{v}{s}-\binom{v-1}{s}=\binom{v-1}{s-1}=\Delta\).
Moreover, since \(\ell\leq k\) and \(\sum_i|X_i|\geq q\),
we have \(ku+\ell(q-1)\leq k(u+q-1)=k(v-\sum_i|X_i|+q-1)\leq k(v-1)\).
By Lemma~\ref{lem:count-vertex-subpartition},
the corresponding rank certificate has value
\(c+\sum_{i=1}^q(k|X_i|-\ell)=R+c-ku-\ell(q-1)\geq R+\Delta-k(v-1)\geq R+\mu\).
The empty certificate has value \(m\), which also satisfies \(m\geq R+\mu\),
since \(m-(R+\mu)=m-\Delta+\ell=\binom{v-1}{s}+\ell\geq0\).

Choose $v$ large enough that $\mu\ge0$. The full-block certificate $\{V\}$ has value $R$, proving $r(M)=R$. Deleting at most $\mu$ edges decreases any other certificate by at most $\mu$, while the full-block value stays $R$. Thus $\lambda(M)>\mu$. Deleting the $\Delta$ edges at one vertex leaves rank at most $k(v-1)-\ell<R$, giving the upper bound.

Averaging a base under the vertex permutations gives a fractional base packing of total weight $m/R$, and the empty flat gives the reverse inequality. Since $\Delta/m=s/v$ and $kv/\Delta\to0$ for $s\ge3$, the ratio tends to $ks$.
\end{proof}

For the graph lower bound, take $v>2k$ and put $M:=M_{k,\ell}(K_v)$. If $D\subseteq E(K_v)$ has $|D|<v-2k$, every nontrivial cut of $K_v-D$ has size at least $v-1-|D|\ge2k$. Nash--Williams--Tutte~\cite{Nash-Williams_1961,Tutte_1961} gives $k$ edge-disjoint spanning trees. Minimum degree at least $2k$ gives at least $kv$ edges, leaving at least $k$ edges outside these trees. Add one distinct unused edge to each of $k-\ell$ trees. The resulting $\ell$ trees and $k-\ell$ spanning unicyclic graphs form an independent set of size $kv-\ell$ by \eqref{eq:count-union}. Thus $r(M)=kv-\ell$ and $\lambda(M)\ge v-2k$. Since the empty flat gives $\sigma(M)\le\binom v2/(kv-\ell)$,
\[
\cogirthratio(M)\ge\frac{(v-2k)(kv-\ell)}{\binom v2}=2k-O(1/v).
\]
For $s\ge3$, Proposition~\ref{prop:complete-uniform-count-ratio} gives the lower bound $ks$.

The LP gap bound $ks$ is sharp for every fixed $p$. Let $M_v$ be the complete-uniform examples above, using complete graphs when $s=2$, and add $p-1$ coloops to obtain $N_v$. Adding vertex-disjoint single-edge components realizes $N_v$ as the count matroid of a hypergraph with edge size at most $s$. Averaging a base under vertex permutations gives $\sigma(M_v)=|E(M_v)|/r(M_v)\to\infty$, while $\cogirthratio(M_v)\to ks$. Every rank-$p$ reducing set in $N_v$ lowers the rank of $M_v$, and deleting all added coloops together with a minimum cocircuit is feasible. Hence
\[
\lambda(M_v)\le\lambda_p(N_v)\le\lambda(M_v)+p-1.
\]
Also $\lambda_p^*(N_v)=\sigma(M_v)+p-1$. Indeed, assigning one to the coloops and using an optimal rank-one solution on $M_v$ proves the upper bound. If a feasible solution assigns total $t\le p-1$ to the coloops, its other coordinates meet each base of $M_v$ to extent at least $p-t$. Its total cost is therefore at least $(p-t)\sigma(M_v)+t\ge\sigma(M_v)+p-1$, since $\sigma(M_v)\ge1$. Dividing the displayed bounds by $\sigma(M_v)+p-1$ proves that $\lambda_p(N_v)/\lambda_p^*(N_v)\to ks$.

\end{document}